\documentclass[runningheads]{llncs}
\usepackage{graphicx}
\usepackage{mwe} 
\usepackage{wrapfig}
\usepackage{caption}
\usepackage[normalem]{ulem}
\usepackage{graphicx,subcaption,ragged2e}
\usepackage{enumitem}
\usepackage[misc]{ifsym}
\usepackage{authblk}
     
\begin{document}

\title{Self-Supervision and Multi-Task Learning: Challenges in Fine-Grained COVID-19 Multi-Class Classification from Chest X-rays}

\titlerunning{COVID-19 Multi-Class Classification from Chest X-rays}

\author{Muhammad Ridzuan$^{(\textrm{\Letter})}$ \and
Ameera Bawazir \and
Ivo Gollini Navarrete \and
Ibrahim Almakky \and
Mohammad Yaqub}

\authorrunning{M. Ridzuan et al.}

\institute{Mohamed bin Zayed University of Artificial Intelligence, Abu Dhabi,\\United Arab Emirates\\
\email{\{muhammad.ridzuan, ameera.bawazir, ivo.navarrete, ibrahim.almakky, mohammad.yaqub\}@mbzuai.ac.ae}}

\maketitle              

\begin{abstract}
Quick and accurate diagnosis is of paramount importance to mitigate the effects of COVID-19 infection, particularly for severe cases. Enormous effort has been put towards developing deep learning methods to classify and detect COVID-19 infections from chest radiography images. However, recently some questions have been raised surrounding the clinical viability and effectiveness of such methods. In this work, we investigate the impact of multi-task learning (classification and segmentation) on the ability of CNNs to differentiate between various appearances of COVID-19 infections in the lung. We also employ self-supervised pre-training approaches, namely MoCo and inpainting-CXR, to eliminate the dependence on expensive ground truth annotations for COVID-19 classification. Finally, we conduct a critical evaluation of the models to assess their deploy-readiness and provide insights into the difficulties of fine-grained COVID-19 multi-class classification from chest X-rays.
\keywords{COVID-19 \and X-ray \and classification \and self-supervision \and multi-task learning.}
\end{abstract}

\section{Introduction}

On January 30, 2020, the World Health Organization (WHO) declared a global health emergency due to the \emph{coronavirus disease 2019} (COVID-19) outbreak \cite{BURKI2020292}. Five times more deadly than the flu, SARS-CoV-2 viral infection's main symptoms are fever, cough, shortness of breath, and loss or change of smell and taste \cite{Struyf20}. The fast-paced rise in infections and the rate at which it spread around the globe exposed many challenges with diagnosis and treatment. Access to screening strategies and treatment was minimal due to the lack of resources, especially at the start of the pandemic \cite{coccolini2021pandemic}.

Polymerase Chain Reaction (PCR) became the gold-standard method for COVID-19 screening. Nevertheless, the limited number of tests and high rate of false negatives (100\% false negative on day one of infection and 38\% on day 5) gave radiographers grounds to define chest imaging not just as a routine screening standard, but as an integral tool for assessing complications and disease progression \cite{inui2021role}. Chest imaging is especially necessary for symptomatic patients that develop pneumonia, which is characterized by an increase in lung density due to inflammation and fluid in the lungs \cite{Cleverley2020}. The Radiological Society of North America (RSNA) developed a standard nomenclature for imaging classification of COVID-19 pneumonia composed by four categories: negative for pneumonia, typical appearance, indeterminate appearance, and atypical appearance of COVID-19 pneumonia \cite{simpson2020radiological}. 

The presence of ground glass opacities (GGOs) and the extent to which they cover lung regions allow radiologists to diagnose COVID-19 pneumonia in chest radiographs. In such manner, the RSNA classifies a case as ``typical'' if the GGOs are multifocal, round-shaped, present in both lungs, and peripheral with a lower lung-predominant distribution. In an ``indeterminate'' case, there is an absence of typical findings and the GGOs are unilateral with a predominant distribution in the center or upper sections of the lung. If no GGO is seen and another cause of pneumonia (i.e. pneumothorax, pleural effusion, pulmonary edema, lobar consolidation, solitary lung nodule or mass, diffuse tiny nodules, cavity) is present, the case is categorized as ``atypical'' \cite{litmanovich2020review}. However, the distinction between these classes is a non-trivial task due to the lack of visual cues and the nature of GGOs, as discussed later in this work.

The possibility of using artificial intelligence (AI) to aid in the fight against COVID-19 motivated researchers to turn to deep learning approaches, especially convolutional neural networks (CNNs), for the detection and classification of COVID-19 infections \cite{Alghamdi}. Many studies have reported high classification performances using chest X-ray radiographies (CXRs) \cite{ji2021research,pham2021classification} and computed tomography (CT) scans \cite{barstugan2020coronavirus,pathak2020deep,jia2021classification} using standard and off-the-shelf CNNs. Despite high reported accuracies by these methods, questions have been raised regarding their clinical usefulness due to the bias of small datasets, poor integration of multistream data, variability of international sources, difficulty of prognosis, and the lack of collaborative work between clinicians and data analysts \cite{roberts2021common}. 

In this investigation, we utilize a large, multi-sourced chest X-ray dataset of COVID-19 patients to train deep learning models that have proven effective on computer vision benchmarks. Following this, the models are evaluated and the results are analysed to identify potential weaknesses in the models. The nature of the data and classes are also analysed keeping in mind the clinical needs for the development of such models. Finally, we present an in-depth discussion into the main challenges associated with this task from the machine learning and data perspectives. Our work does not aim to outperform state-of-the-art publications, but rather provide important insights in this challenging problem. The contributions of this work are summarized as follows:
\begin{enumerate}[topsep=1pt,itemsep=1ex,partopsep=1ex]
    \item Investigating the impact of multi-task learning (classification and segmentation) on the ability of CNNs to differentiate between various appearances of COVID-19 infections in the lung.
    \item Employing self-supervised pre-training approaches, namely MoCo and inpaint-ing-CXR, to eliminate the dependence on expensive ground truth annotations for COVID-19 classification.
    \item Conducting a critical evaluation of the best performing model to provide insights into the difficulties of fine-grained COVID-19 multi-class classification from chest X-rays.

\end{enumerate}

\section{Method}
 In this work, we trained different deep model architectures to classify each input X-ray image into one of four classes: negative for pneumonia, typical, indeterminate, or atypical appearance of COVID-19. We employed two main elements of the SIIM-FISABIO-RSNA COVID-19 \cite{challenge_dataset} winning solutions, namely pre-training and multi-task learning, and compared them against self-supervised learning approaches to analyze their impact and generalizability on fine-grained COVID-19 multi-class classification.

\subsection{Baseline model}
\label{section:baseline}
The baseline CNN was chosen from four architectures to explore the performance of lightweight models like MobileNet \cite{howard2017mobilenets} and EfficientNet \cite{tan2020efficientnet} against dense models such as ResNet \cite{he2015deep} and DenseNet \cite{huang2018densely}. DenseNet-121 was selected for comparison and evaluation of the different approaches due to its balance between accuracy and training speed, supported by its success with CXRs reported in literature \cite{chexnet_rajpurkar}. 

\subsection{Multi-task learning}
Multi-task learning aims to utilize information from multiple tasks in order to improve the performance and generalizability of the model \cite{Zhang2021}. Inspired by the winning solution of the SIIM-FISABIO-RSNA COVID-19 Detection challenge \cite{siim_1stplace}, we employ multi-task learning to improve the classification performance. The first stage of the solution consists of a pre-trained encoder for classification, while the second stage consists of a pre-trained encoder-decoder architecture that is later fine-tuned on the COVID-19 dataset to learn both classification and segmentation tasks. The segmentation was performed using the ground truth bounding boxes converted to opacity masks. A Dice-Weighted Cross Entropy loss is used as the multi-task loss, where $W_{CE}$ and $W_{Dice}$ are the weights for each loss component, and $w_{c}$ is the weight of each class:
\begin{equation}\label{eq:loss}
L_{DiceWCE} = W_{CE} (- \sum_{c=1}^{C} w_c y^c \log \hat{y}^c) + W_{Dice} (1-\frac{2\sum_{c=1}^C \sum_{i=1}^{N} g_i^c s_i^c}{\sum_{c=1}^C \sum_{i=1}^{N} g_i^c + \sum_{c=1}^C \sum_{i=1}^{N} s_i^c})
\end{equation}

\subsection{Self-supervised pre-training}
Self-supervised pre-training has proven effective in numerous medical tasks with a scarcity of labelled training data, e.g. \cite{SSL_ex1,SSL_ex2,sowrirajan2021mococxr}. Self-supervised deep CNN models have also been employed to classify COVID-19 cases from chest X-ray images and to deal with the problem of class imbalance \cite{gazda2021self}. We employed MoCo \cite{sowrirajan2021mococxr} and inpainting \cite{inpainting2016} as constrastive and generative self-supervised learning (SSL) techniques trained on a large unlabelled dataset, then fine-tuned on the COVID-19 dataset to classify the above-mentioned four classes. 

\subsubsection{MoCo-CXR.} Adding to the work of \cite{sowrirajan2021mococxr}, we introduced further augmentation strategies to the MoCo-CXR architecture: horizontal translation, random scaling, and rotation. We also decreased the InfoNCE loss \cite{infoNCE_loss} temperature value, where InfoNCE is defined as
\begin{equation}
L_q = -log (\frac{exp(qk_+ / \tau)}{\sum_{i=0}^k exp(qk_i /\tau)})
\end{equation}
where it measures the cosine distance between a key $k_{i}$ and query input $q$ and applies scaling to the distance by a temperature parameter $\tau$. We pre-trained DenseNet-121 using the original MoCo-CXR \cite{sowrirajan2021mococxr}, the modified MoCo-CXR approach, and MoCo-V2 \cite{chen2020improved}.

\subsubsection{Inpainting-CXR.} 
\begin{figure}[tp]
    \centering
    \includegraphics[scale=0.47]{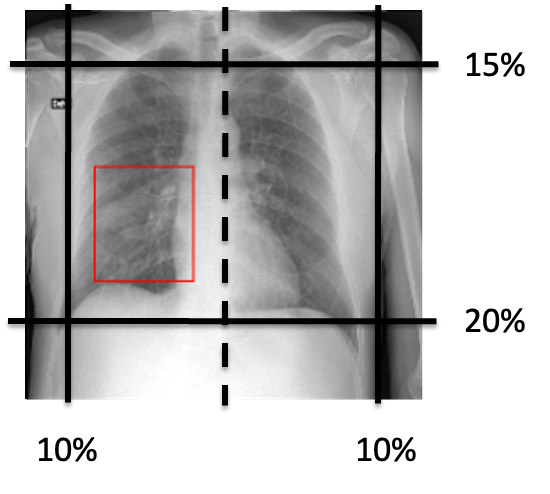}
    \caption{Mask region constraint for SSL Inpainting-CXR. This is performed by posing the following constraints: 10\% from the left and right, 15\% from the top, and 20\% from the bottom of the chest X-rays.}
    \label{fig:mask_constraint}
    
\end{figure}

\begin{figure}[t]
\centering
  \begin{tabular}[c]{c}
     \includegraphics[scale=0.5]{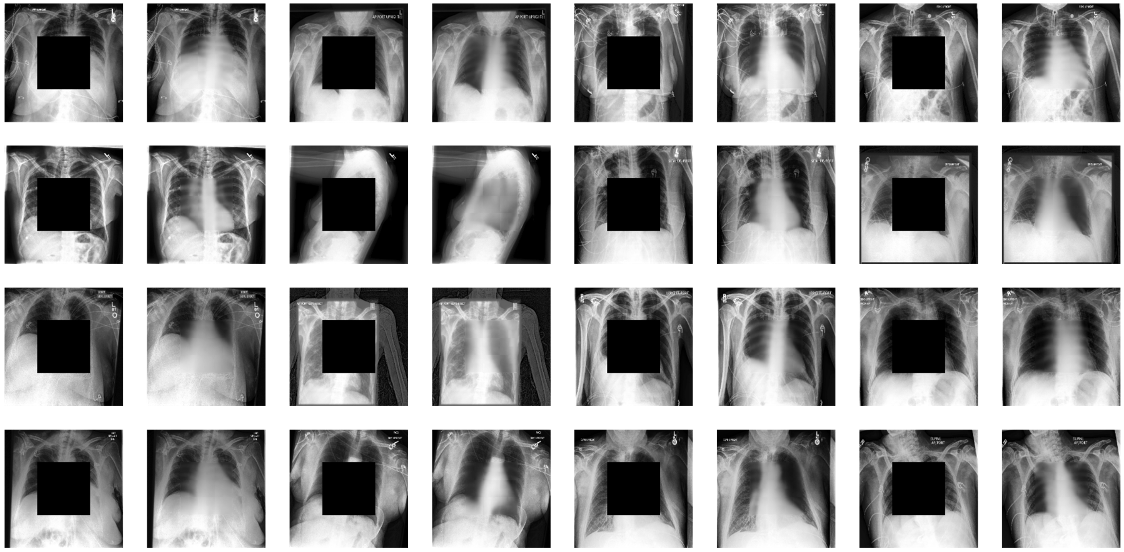} \\
    \small (a)
  \end{tabular}
  \hspace{1em}
  \begin{tabular}[c]{c}
     \includegraphics[scale=0.5]{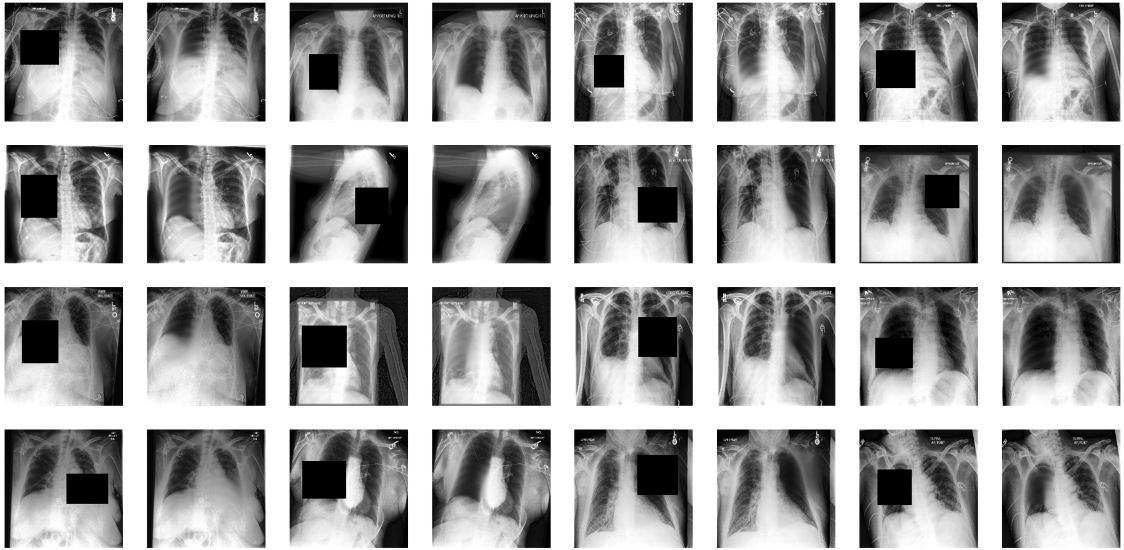}\\
    \small (b)
  \end{tabular}
  \caption{Visualization of inpainting self-supervised pre-training model output, showing image reconstruction from (a) center mask and (b) targeted left and right mask (inpainting-CXR).}
\label{fig:inpainting_masks}
\end{figure}

Based on \cite{inpainting2016}, we also explored the impact of Inpainting-CXR, a focused lung masking inpainting strategy on the model's ability to learn effective representations to identify chest abnormalities. Using this as a pretext task, we applied targeted lung masking by approximating its location for both lungs (Figure \ref{fig:mask_constraint}).  Additionally, center inpainting was also explored, where a center mask is created on the X-ray images, and the model is tasked with reconstructing the original masked region. Figure \ref{fig:inpainting_masks} shows the center mask and Inpainting-CXR (left and right targeted lung mask) with the reconstructed images. 

\section{Dataset}
\label{section:dataset}
The primary dataset used for classification in this study is the SIIM-FISABIO-RSNA COVID-19 Detection dataset curated by an international group of 22 radiologists \cite{challenge_dataset}. It includes data from the Valencian Region Medical ImageBank (BIMCV) \cite{vaya2020bimcv} and the Medical Imaging Data Resource Center (MIDRC) - RSNA International COVID-19 Open Radiology Database (RICORD) \cite{tsai2021rsna}. The dataset available for training is composed of 6,054 individual cases (6,334  radiographs), with each case being labelled as negative, typical, indeterminate, or atypical appearance of pneumonia. The dataset also includes radiologists' bounding box annotations that highlight regions of opacities within the chest X-rays.

In addition to the SIIM-FISABIO-RSNA dataset, the CheXpert \cite{irvin2019chexpert} and RSNA \cite{kaggle_rsna} datasets were also used for pre-training. CheXpert is a multilabel dataset composed of 224,316 chest radiographs of 65,240 patients with the presence of 13 chest abnormalities. RSNA is a subset of CheXpert; a pneumonia detection dataset consisting of CXRs from 26,684 patients labelled according to the presence or absence of lung opacities.
With such large number of chest radiographs containing various manifestations in the lungs including pneumonia, pleural effusion, consolidation and others, these datasets were chosen for pre-training.

\section{Experimental Settings}
\label{section:experimental_settings}
The baseline DenseNet-121 CNN model was trained with an input size of $224 \times 224$ and batch size of 16. A cross-entropy loss function along with ADAM optimizer and learning rate of $0.001$ were used. The following augmentations were performed: horizontal flip, rotation up to $\pm10$ degrees, and scaling up to $20\%$. For pre-processing, we applied winsorization at $92.5$-percentile and normalization between $0$ and $1$ for the image pixel values. 

The multi-task learning encoder-decoder architecture consists of a UNet with a DenseNet-121 backbone. In the first stage, DenseNet-121 was pre-trained for $30$ epochs on CheXpert \cite{irvin2019chexpert}, then the weights were transferred to initialize the encoder of UNet in the second stage. The encoder-decoder model was trained for segmentation and classification for $30$ epochs on RSNA dataset \cite{kaggle_rsna} using a Dice-Cross Entropy compound loss with equal weights. Finally, the pre-trained weights were used to fine-tune the encoder-decoder on the COVID-19 dataset using a Dice-Weighted Cross Entropy loss with a weight of $0.6$ and $0.4$, respectively. The different classes in Weighted Cross Entropy loss were weighted by $0.2$, $0.2$, $0.3$, and $0.3$ for negative, typical, indeterminate, and atypical respectively. The model was optimized with ADAM optimizer using an initial learning rate of $0.0001$ and Cosine Annealing LR scheduler.         

For self-supervised learning, CheXpert \cite{irvin2019chexpert} was used without labels for pre-training. DenseNet-121 was chosen as the key and queue encoder for MoCo-CXR, modified MoCo-CXR, and MoCo-V2. For MoCo-V2, the model applied the default augmentations from \cite{sowrirajan2021mococxr} which included $224\times224$ pixel crop, random color jittering, random horizontal flip, random gray scale conversion, and image blur. We evaluated MoCo-CXR \cite{chen2020improved} with the two default augmentations, horizontal flipping and random rotation of $\pm10 $ degrees. We further modified the MoCo-CXR augmentation strategy by increasing the rotation angle to $\pm20$  degrees, adding horizontal translation $20\%$, randomly scaling the image in range of $100-120\%$, and decreasing the temperature value of the InfoNCE loss from $0.2$ to $0.07$.

For inpaiting SSL, we substituted the original AlexNet encoder with DenseNet-121. We trained the model using the Mean Squared Error (MSE) loss, and omitted the adversarial loss to focus on transferability rather than fine reconstruction. The targeted left and right lung masking (inpainting-CXR) was used with varying mask sizes from $17\times17$ up to $32\times32$, while the center inpainting used mask of size $100\times100$. 

\section{Results and Discussion}
We first chose an appropriate baseline, then investigated the impact of multi-task classification and segmentation learning, followed by self-supervised learning. Tables \ref{table:experimental_results} and \ref{table:experimental_results_5fold} summarize our experimental results on the SIIM-FISABIO-RSNA dataset. Assessing the best performing models from the 80/20\% split (Table \ref{table:experimental_results}) on the 5-fold cross-validation set (Table \ref{table:experimental_results_5fold}) allows for better judgment surrounding the generalizability of the model. We focus on the $F_1$-macro score to ensure fair comparison between the models considering the unbalanced nature of the classes. Our best performing baseline architecture is DenseNet-121, consistent with its reported success with CXRs in literature \cite{chexnet_rajpurkar} (Table \ref{table:baseline}).

\begin{table}[bp]
\centering
  \caption{Experimental results for 80-20\% train-test split of the SIIM-FISABIO-RSNA dataset.}%
  \begin{tabular}{|p{160pt}|p{60pt}|p{60pt}|}
   \hline
  \bfseries Approach & \bfseries $F_1$ Score & \bfseries Acc. (\%) \\
  \hline\hline
  
    Baseline: DenseNet-121 & 0.4205 & 57.34 \\
    \hline\hline
    Multi-task (MT) without pre-training & 0.4300 & 64.16\\
    \hline
    MT pre-trained RSNA enc. & 0.4401 & 62.23\\
    \hline
    MT pre-trained RSNA enc. + dec. & \textbf{0.4794} & \textbf{61.77}\\
    \hline\hline
    SSL MoCo-CXR  & 0.4305 & 58.69\\
    \hline
    SSL Modified MoCo-CXR  & 0.4325 & 59.71 \\
    \hline
    SSL MoCo-V2  & 0.4583 & 58.60 \\
    \hline
    SSL Center inpainting & 0.4472 & 59.62 \\
    \hline
    SSL Inpainting-CXR & \textbf{0.4706} & \textbf{65.22}\\
    \hline
  
  \end{tabular}
  
\label{table:experimental_results}
\end{table}

\begin{table}
\centering
  \caption{Experimental results for 5-fold cross-validation on the baseline and top performing solutions of the SIIM-FISABIO-RSNA dataset.}%
  \begin{tabular}{|p{140pt}|p{80pt}|p{80pt}|}
   \hline
  \bfseries Approach & \bfseries $F_1$ Score & \bfseries Acc. (\%) \\
  \hline\hline
  
    Baseline: DenseNet-121 & 0.4205 $\pm$ 0.0149 & 57.34  $\pm$ 2.79 \\
    \hline
    MT pre-trained RSNA enc. + dec. & \textbf{0.4621}  $\pm$ \textbf{0.0099} & \textbf{64.07} $\pm$ \textbf{1.21} \\
    \hline
    SSL Inpainting-CXR & \textbf{0.4599 $\pm$ 0.0137} & \textbf{59.55 $\pm$ 1.07}\\
    \hline
  \end{tabular}
\label{table:experimental_results_5fold}
\end{table}

\begin{table}
\centering
  \caption{Comparison of CNN architectures to define baseline.}%
  \begin{tabular}{|p{140pt}|p{60pt}|p{60pt}|}
  \hline
  \bfseries Experiments & \bfseries $F_1$ Score & \bfseries Acc. (\%)  \\
  \hline
  MobileNet & 0.3356 & 57.18 \\
  \hline
  EfficientNet & 0.3434 & 61.47 \\
  \hline
  ResNet-50 & 0.1617 & 47.80\\
  \hline
  DenseNet-121 & \bfseries 0.4345 & 58.19\\
  \hline
  \end{tabular}
\label{table:baseline}
\end{table}

Following the baseline establishment, we evaluated the impact of multi-task segmentation and classification learning on COVID-19 fine-grained classification. Tables \ref{table:experimental_results} and \ref{table:experimental_results_5fold} show the improvement gain from incorporating a segmentation head and using RSNA pre-training, providing evidence for the effectiveness of including bounding box annotations during training to localize the model's learning on the regions of infection. Notably, our self-supervised approaches also achieve comparable performance, with inpainting-CXR outperforming MoCo-CXR. We hypothesize that the superiority of inpainting-CXR over MoCo-CXR is likely due to its generative nature where the model is forced to learn the distinctive features of the lungs and diseases through reconstruction. While the performance of inpainting-CXR is slightly lower on average than multi-task learning, this method completely avoids the need for expensive bounding box annotations, highlighting an exciting avenue for further exploration particularly in the use of generative self-supervised learning for CXR pathology identification.
\par Nevertheless, a question remains regarding the deploy-readiness of current models in clinical settings. In the following sections, we provide a critical evaluation of the solutions and demonstrate the challenges and difficulties of COVID-19 fine-grained classification. For further evaluation of the results, we present a thorough analysis with regards to model interpretability.

\subsection{Lack of visual cues in COVID-19 CXRs} 
\begin{figure}[htp]
    \centering
    \includegraphics[scale=0.55]{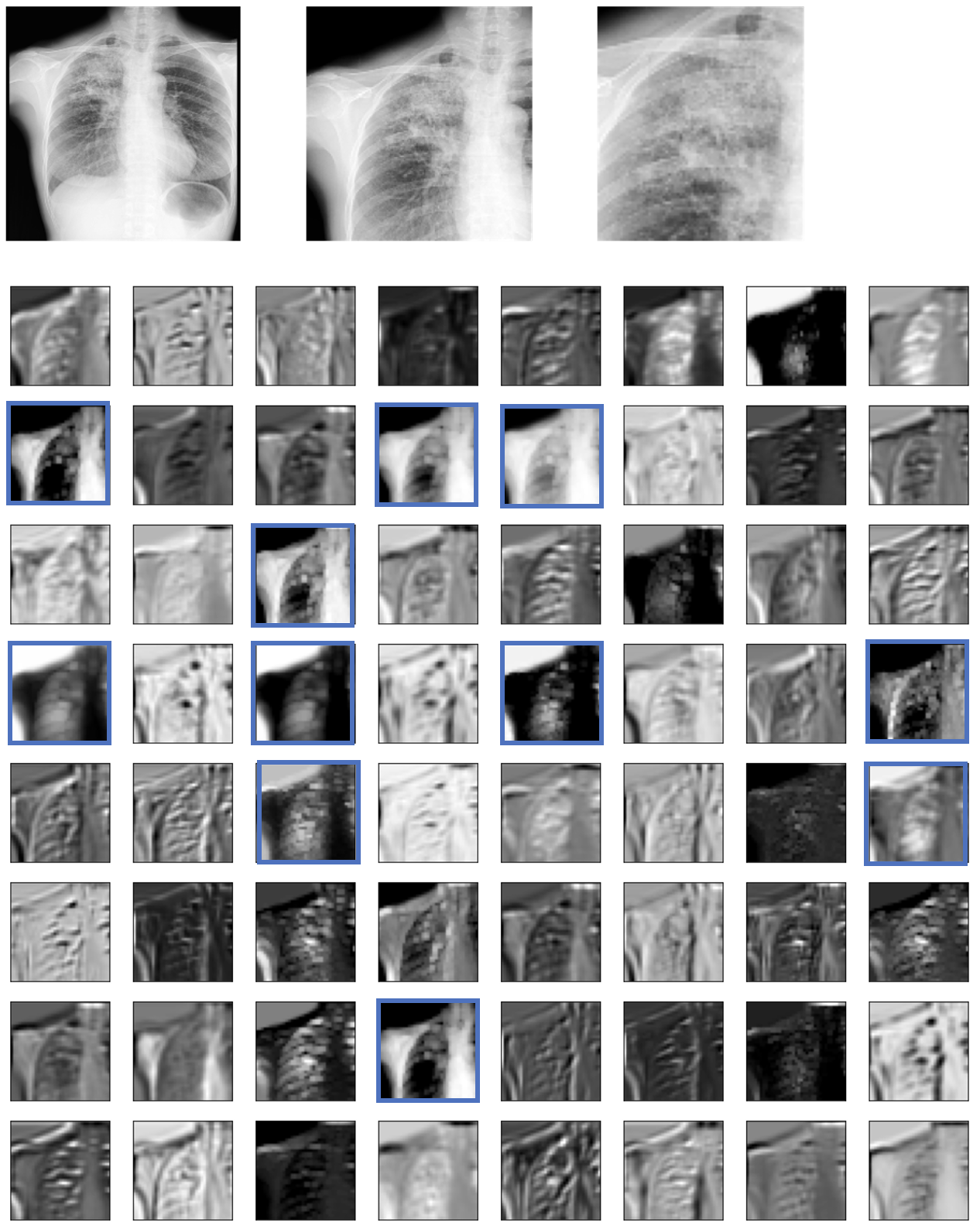}
    \caption{Tuberculosis feature maps. The top three images correspond to the original CXR (image source \cite{tb_liu2020}), a zoomed-in section nearing the area of infection from which the feature maps are created, and a zoomed-in section directly on the area of infection based on the radiologist’s ground truth annotation. We outlined some feature maps that we think may have been pertinent to the model’s ability to delineate the features of tuberculosis in the upper lobe of the right lung. The feature maps were created using DenseNet-121 pre-trained on CheXpert and fine-tuned on TBX11K \cite{tb_liu2020}.}
    \label{fig:TB_fea_map}
\end{figure}

\begin{figure}[htp]
    \centering
    \includegraphics[scale=0.55]{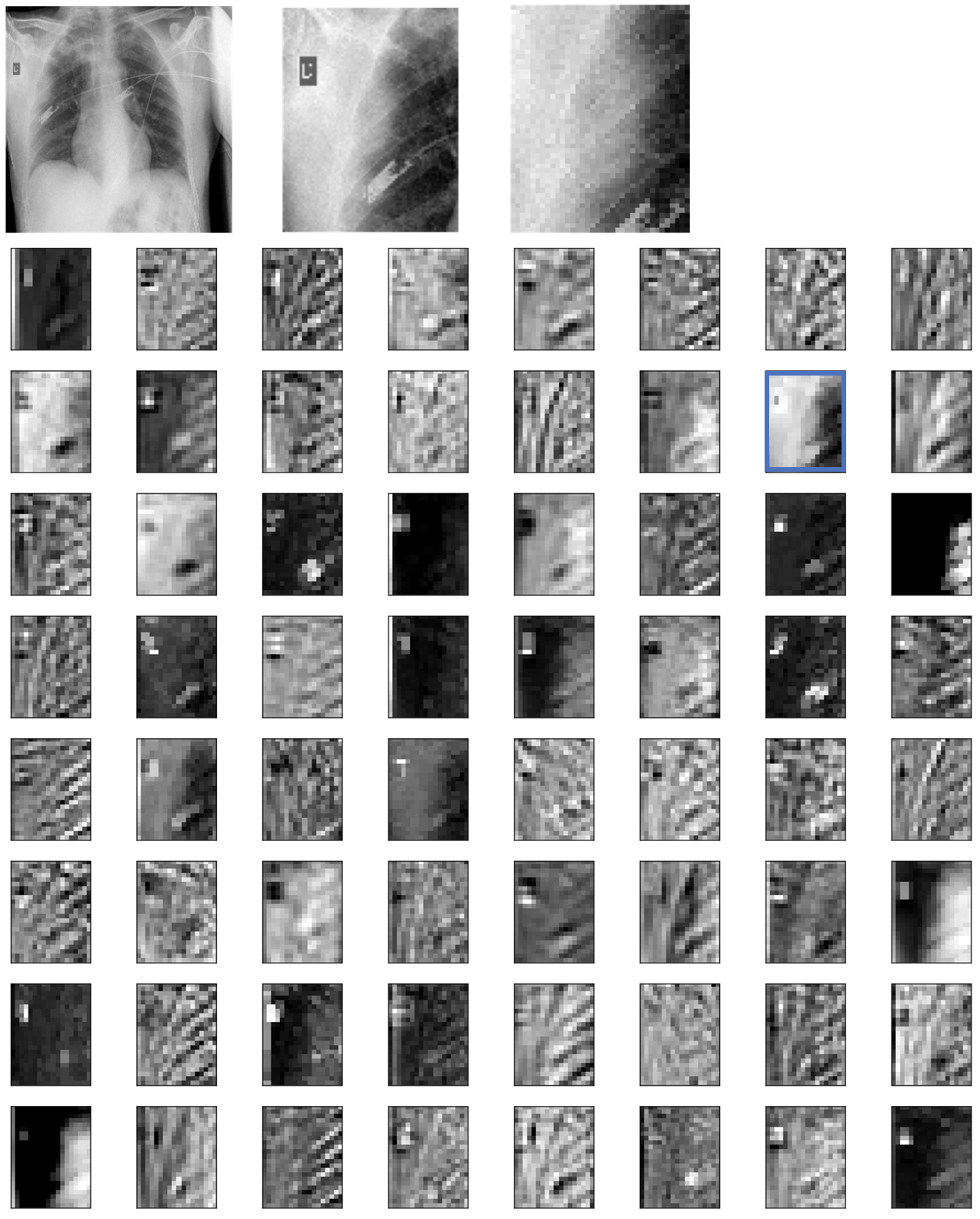}
    \caption{COVID-19 feature maps. The top three images correspond to the original CXR, a zoomed-in section nearing the area of infection from which the feature maps are created, and a zoomed-in section directly on the area of infection based on the radiologist’s ground truth annotation. We outlined the feature map that we think may have been pertinent to the model’s ability to delineate the features of COVID-19. Most feature maps were unable to identify the features of infection on the periphery of the left lung. The feature maps were created using DenseNet-121 pre-trained on CheXpert and fine-tuned on SIIM-FISABIO-RSNA COVID-19 \cite{challenge_dataset}.}
    \label{fig:COVID19_fea_map}
\end{figure}

CNNs have recorded successes in many medical applications, including in CXRs. Unlike the manifestations of other pathologies such as cardiomegaly (which is characterized by the size of the object of interest) or tuberculosis (which is characterized by the presence of bacteria occurring in colonies), the precise identification of COVID-19 from CXRs presents a particular challenge even to the experienced radiologists \cite{williams2013variability}. We present the feature maps of tuberculosis and COVID-19 in Figures \ref{fig:TB_fea_map} and \ref{fig:COVID19_fea_map} respectively, and show that the learned representations of the latter is much harder than the former. This is likely due to the prevalence of ground-glass opacities in COVID-19 which tend to be more dispersed, fine-to-absent in texture, and typically less concentrated in appearance \cite{radCTperformance,cozzi2021ground,Hansell2008FleischnerSG} -- that even on images with more obvious manifestations of COVID-19 (as in Figure \ref{fig:COVID19_fea_map}), the model is unable to learn proper representations. From a machine learning perspective, the challenge for CNNs stems from the lack of well-defined shapes, sizes, or edges of the infections. CNNs, on the other hand, heavily rely on edge detections, as evidenced by the feature maps shown.

\begin{figure}[htp]
    \centering
    \includegraphics[scale=0.43]{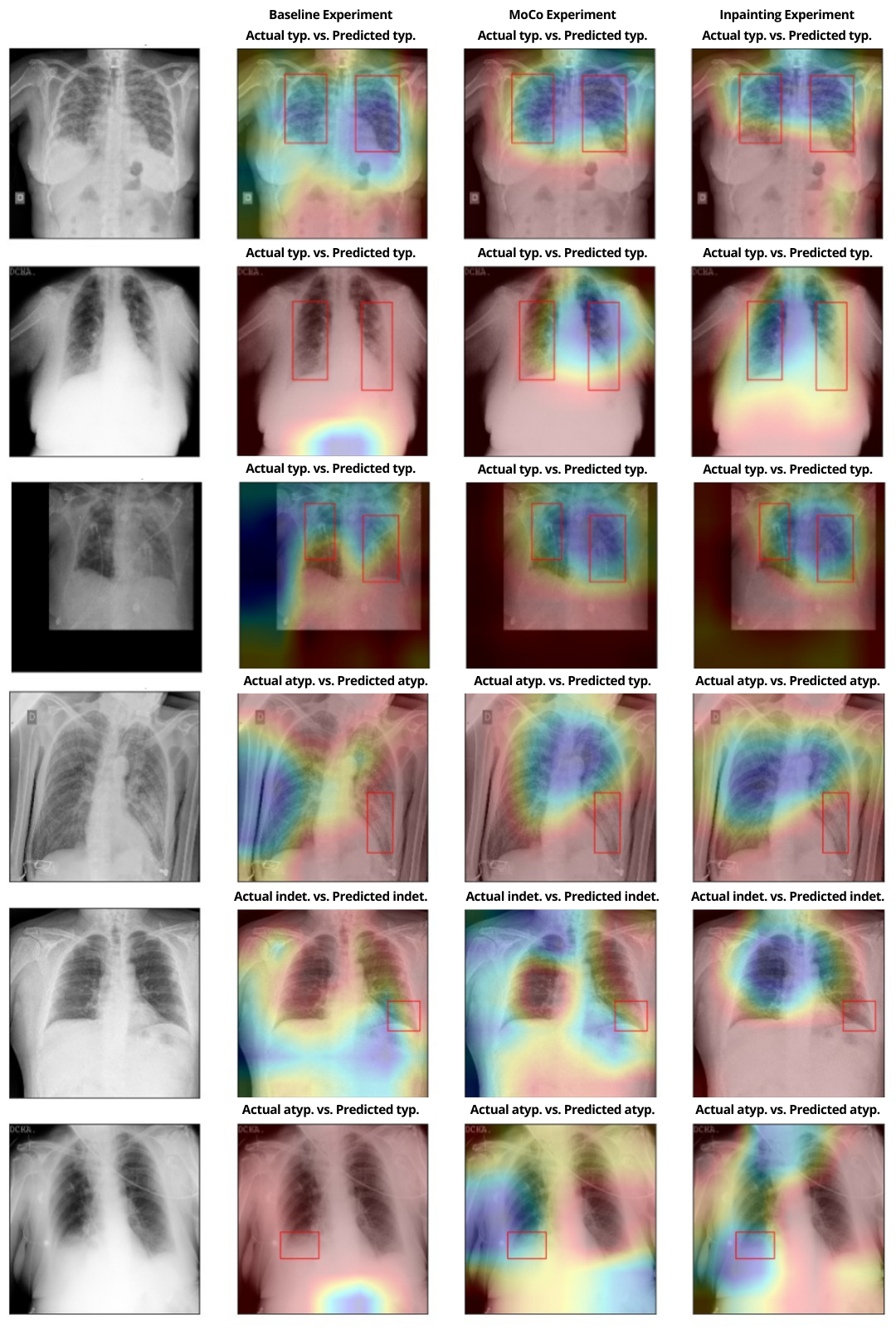}
    \caption{Successes and failures of GradCAM. The top images (rows 1-3) show examples where the GradCAM outputs correctly localize the regions of infection, while the bottom images (rows 4-6) show examples of incorrect predictions. The bounding boxes display the ground truth radiologists’ annotations as provided by the challenge. The heatmap importance increases from red to blue.}
    \label{fig:gcam_full}
\end{figure}

\subsection{(Mis)learned representations} 
As we have demonstrated the difficulty of COVID-19 detection from CXRs, the question then becomes, what does the model learn? We present the GradCAM \cite{gradcam} outputs of the baseline, MoCo, and inpainting experiments in Figure \ref{fig:gcam_full}. Evidently, the model is better able to localize the lung regions when using MoCo and inpainting self-supervised pre-training (Figure \ref{fig:gcam_full}; columns 3 and 4), while the decision-making appears to be more sporadic and irregular on the baseline model (Figure \ref{fig:gcam_full}; column 2). 

\begin{figure}[tp]
    \centering
    \includegraphics[scale=0.4]{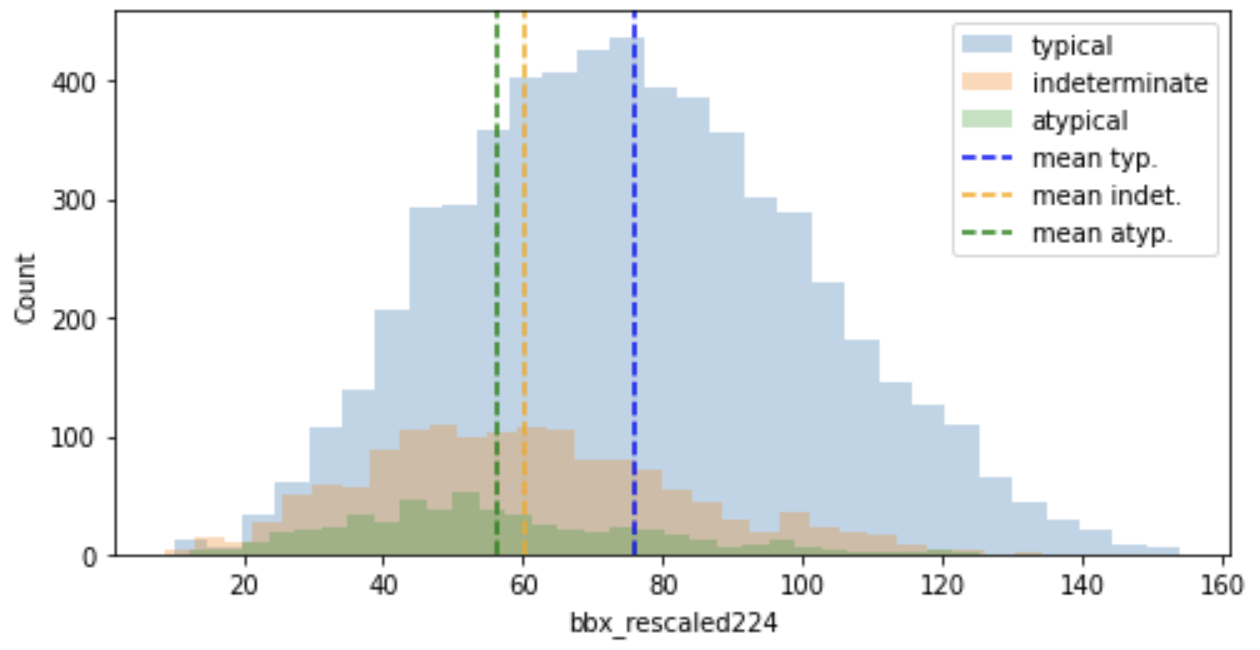}
    \caption{Approximate distribution of the ground truth bounding box dimensions after re-scaling the CXRs to $224\times224$, generated by taking the square root of the bounding box areas. The mean bounding box dimension of the typical class is higher than that of atypical and indeterminate.}
    \label{fig:bbx_size_distribution}
\end{figure}

A closer inspection of the SSL GradCAMs reveals that the correctly predicted regions (Figure \ref{fig:gcam_full}; rows 1-3) are extensive, encompass the majority of the lungs, and have large corresponding regions of infection. We note that the correctly predicted images also belong primarily to the typical class, and that the average areas of infection for the typical class, represented by the bounding box dimensions, is higher than that of the atypical and indeterminate classes (Figure \ref{fig:bbx_size_distribution}). Conversely, the incorrectly predicted regions (Figure \ref{fig:gcam_full}; rows 4-6) are also extensive and highlight the majority of the lungs, but have smaller regions of infection that are often missed by the model. This suggests that the correct predictions of the model may not necessarily be attributed to the true regions of infection within the lungs, but rather to some other non-causal or false positive features of the lung. 

\subsection{COVID-19 binary classifications}
\begin{figure}
\captionsetup[subfigure]{justification=Centering}

\centering
    \begin{subfigure}[t]{0.42\textwidth}
        \includegraphics[width=\textwidth]{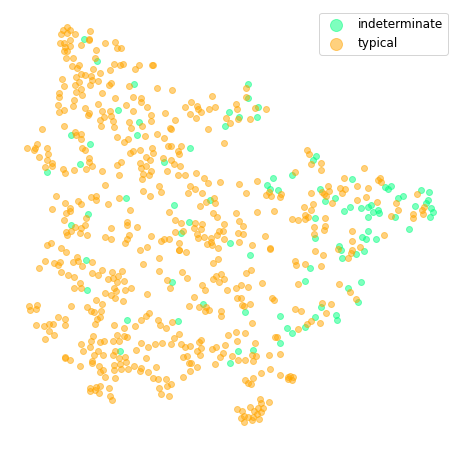}
        \caption{}
    \end{subfigure}\hspace{\fill} 
    \begin{subfigure}[t]{0.42\textwidth}
        \includegraphics[width=\linewidth]{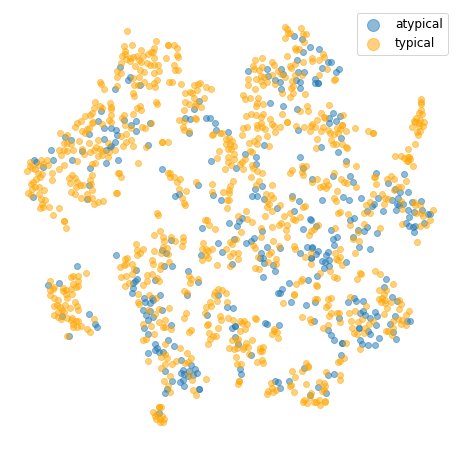}
        \caption{}
    \end{subfigure}
\begin{subfigure}[t]{0.42\textwidth}
    \includegraphics[width=\linewidth]{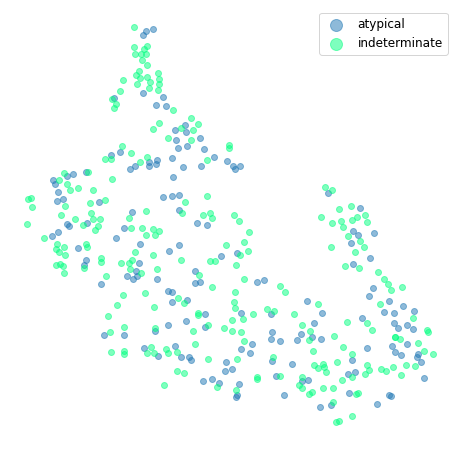}
    \caption{}
\end{subfigure}\hspace{\fill} 
\begin{subfigure}[t]{0.42\textwidth}
    \includegraphics[width=\linewidth]{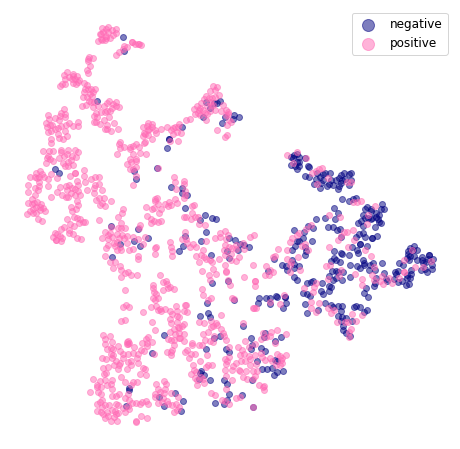}
    \caption{}
\end{subfigure}

\caption{t-SNEs of the last DenseNet-121 block of our best-performing model conditioned on the true labels for (a) typical-indeterminate, (b) typical-atypical, (c) atypical-indeterminate, and (d) positive-negative COVID-19 classes.}
\label{fig:tsne}
\end{figure}

Some studies have demonstrated great successes in the binary classification of COVID-19. For example, \cite{Yang2021_binary_covid19_posneg} achieved a 0.985 $F_1$-score in classifying between positive and negative cases of COVID-19; \cite{Yoo2020_binary_covid19_vs_tb} reported a 1.00-$F_1$-score in distinguishing between COVID-19 and tuberculosis. Arguably, the variation between classes is larger in these cases as the studies were performed to separate interclass differences. We are interested in intraclass differences. We perform an ablation study of each pair of positive classes using DenseNet-121 to investigate the model's ability to differentiate between the COVID-19 cases. Table \ref{table:binary_exps} shows that the results of all pairs of positive classes are only slightly better than chance. 
The t-SNEs (Figure \ref{fig:tsne}) also show poorly defined clusters that alludes to the inability of the model to distinguish between the fine-grained COVID-19 classes. However, a clearer separation is evident in the classification of positive-negative cases, consistent with the findings of other studies (e.g., \cite{Yang2021_binary_covid19_posneg}).

\begin{table}
\centering
  \caption{Comparison of binary classifications for COVID-19 pneumonia appearances.}%
  \begin{tabular}{|p{140pt}|p{60pt}|p{60pt}|}
  \hline
  \bfseries Binary classes & \bfseries $F_1$ Score & \bfseries Acc. (\%)  \\
  \hline
  Typical-Indeterminate & 0.6177 & 70.24 \\
  \hline
  
  Atypical-Indeterminate & 0.5760 & 66.77 \\
  \hline
  
  Typical-Atypical & 0.6252 & 85.96 \\
  \hline
  Positive-Negative & 0.7735 & 81.98 \\
  \hline
  \end{tabular}
\label{table:binary_exps}
\end{table}



\section{Conclusion}
In this work, we have shown the merit of our self-supervised inpainting-CXR method that rivals the performance of multi-task segmentation and classification learning in COVID-19 multi-class classification. We discuss the challenges surrounding the fine-grained distinction of COVID-19 appearances and delve deep into the obstacles hindering the use of current models to develop a clinically viable solution for COVID-19 infections from chest radiographs. We believe among the reasons for the low performance of current models are due to the tendency of CNNs to rely on edge detections and the corresponding lack of visual cues prevalent in COVID-19 CXRs due to the presence of diffused ground-glass opacities. Our work may spark further interest in the community in the use of generative self-supervised learning as a cost-effective method to achieve desirable results without the need for expensive labelled bounding box annotations. 

%
%
%
\bibliographystyle{splncs04}
\bibliography{main}

\end{document}


\section{Supplementary}
\appendix

\begin{table}[htp]
\centering
  \caption{Comparison of CNN architectures to define baseline.}%
  \begin{tabular}{|p{140pt}|p{60pt}|p{60pt}|}
  \hline
  \bfseries Experiments & \bfseries $F_1$ Score & \bfseries Acc. (\%)  \\
  \hline
  MobileNet & 0.3356 & 57.18 \\
  \hline
  EfficientNet & 0.3434 & 61.47 \\
  \hline
  ResNet-50 & 0.1617 & 47.80\\
  \hline
  DenseNet-121 & \bfseries 0.4345 & 58.19\\
  \hline
  \end{tabular}
\label{table:baseline}
\end{table}

\begin{table}[htp]
\centering
  \caption{Preprocessing, image size and augmentation ablations.}%
  \begin{tabular}{|p{140pt}|p{60pt}|p{60pt}|}
  \hline
  \bfseries Preprocessing & \bfseries $F_1$ Score & \bfseries Acc. (\%)  \\
  \hline
  Winsorization 97.5\% & 0.3834 & 55.42 \\
  \hline
  Winsorization 95.0\% & 0.3494 & 62.07 \\
  \hline
  Winsorization 92.5\% & \bfseries 0.4308 & 60.89\\
  \hline
  Winsorization 90.0\% & 0.3804 & 62.87\\
  \hline
  Histogram equalization & 0.3877 & 62.47\\
  \hline\hline
  \bfseries Image size & & \\
  \hline
  128$\times$128 & 0.3845 & 62.87 \\
  \hline
  224$\times$224 & \bfseries 0.4308 & 60.89 \\
  \hline
  256$\times$256 & 0.3441 & 60.41\\
  \hline
  512$\times$512 & 0.4174 & 62.31\\
  \hline
  1024$\times$1024 & 0.3596 & 57.64\\
  \hline\hline
  \bfseries Augmentation &  & \\
  \hline
  No augmentation & 0.4308 & 60.89 \\
  \hline
  RandomRotation (-10, 10) & 0.4393 & 58.59 \\
  \hline
  RandomRotation (-10, 10) + \ \ RandomHorizontalFlip (p=0.5) & 0.4414 & 57.96\\
  \hline
  RandomRotation (-10, 10) + RandomHorizontalFlip (p=0.5) + Scale (1.0, 1.2) &  0.4504 & 58.99\\
  \hline
  RandomRotation (-10, 10) + RandomHorizontalFlip (p=0.5) + Scale (1.0, 1.2) + Shear (0.0, 0.1) & \bfseries 0.4573 & 57.48\\
  \hline
  \end{tabular}
\label{table:aug_ablation}
\end{table}

\begin{figure}[htp]
    \centering
    \includegraphics[scale=0.5]{Images/midl_conf_mtx_inpainting.png}
    \caption{Confusion matrix for the best-performing inpainting-CXR model.}
    \label{fig:conf_mtx}
\end{figure}

\begin{figure}[t]
\centering
  \begin{tabular}[c]{c}
     \includegraphics[scale=0.5]{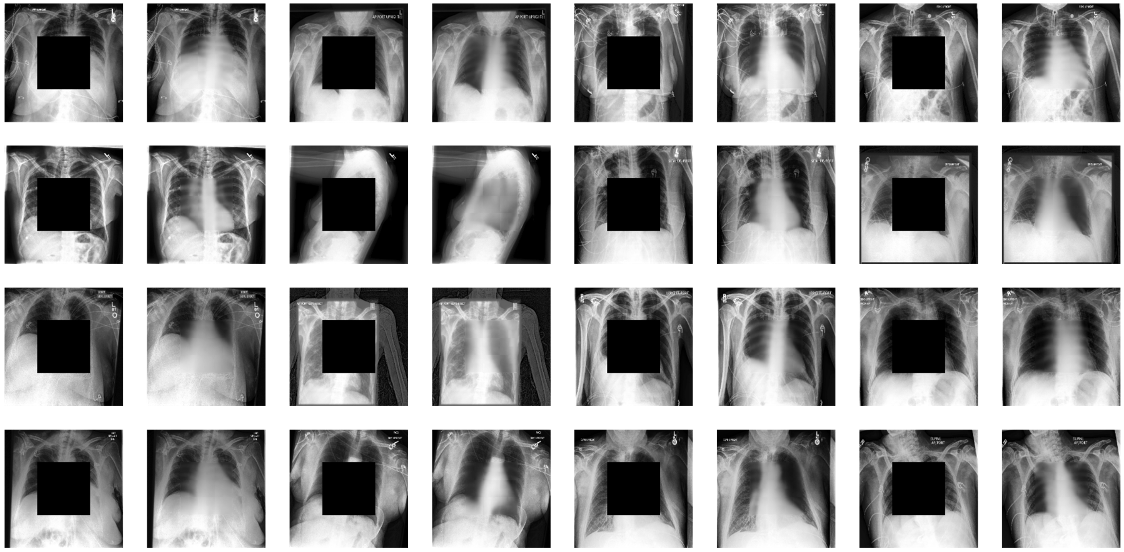} \\
    \small (a)
  \end{tabular}
  \hspace{1em}
  \begin{tabular}[c]{c}
     \includegraphics[scale=0.5]{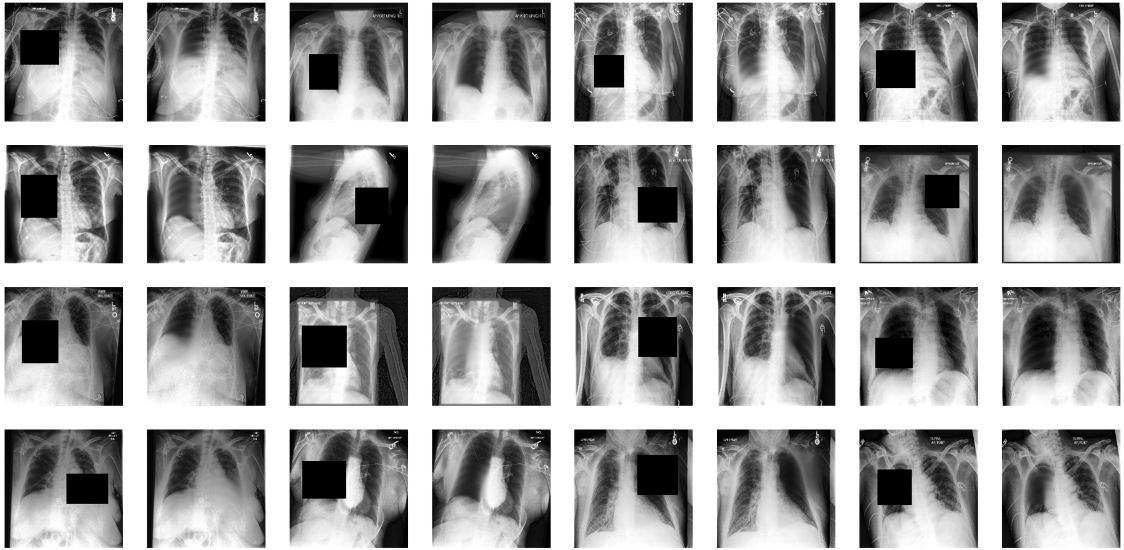}\\
    \small (b)
  \end{tabular}
  \caption{Visualization of inpainting self supervised pre-training model output, showing image reconstruction from (a) center mask and (b) inpainting CXR (targeted left and right mask).}
\label{fig:inpainting_masks}
\end{figure}